\documentclass[a4paper,11pt]{article}
\usepackage{jinstpub} 
\usepackage{subcaption}

\title{\boldmath Performance study of 4-MU-loaded water for Cherenkov light detection}

\author[a]{Pendo B. Nyanda}
\author[b]{Gowoon Kim}
\author[a,b]{Youngduk Kim}
\author[b]{Kyungmin Seo}
\author[b,1]{Jaison Lee,\note{Corresponding author.}}
\author[b]{Olga Gileva}
\author[c]{Eungseok Yi}
\affiliation[a]{University of Science \& Technology,  Daejeon, Korea}
\affiliation[b]{Institute for Basic Science, Daejeon, Korea}
\affiliation[c]{Korean Institute of Geoscience and Mineral Resources, Daejeon, Korea}

\emailAdd{jsahnlee@ibs.re.kr}

\abstract{We report on an R\&D study to improve the photon detection efficiency of water Cherenkov detectors by doping ultrapure water with 4-methylumbelliferone (4-MU), a wavelength-shifting additive. Cherenkov light yields from cosmic-ray muons were measured for various 4-MU concentrations and compared with those from pure water. At a concentration of 1 ppm, the detected light yield increased by approximately a factor of three. This enhancement can be attributed to wavelength shifting and improved photon collection efficiency. No noticeable degradation in optical transparency was observed across the tested concentrations of 0.5 and 1 ppm with different concentration of ethanol. These results suggest that 4-MU is a promising additive for improving the performance of water Cherenkov detectors.\\ 
(Accepted for publication in the Journal of Instrumentation (JINST), 27 January 2026.)}


\arxivnumber{2511.03989} 

\begin{document}
\maketitle
\flushbottom

\section{Introduction}
\label{sec:intro}
Water Cherenkov detectors are widely used in particle and astroparticle physics for their scalability, directional sensitivity, and relatively low cost. Their applications range from neutrino detection to cosmic-ray veto systems ~\cite{FUKUDA2003418, Knapp:2009zz, Geis:2018com}. In such detectors, charged particles traveling faster than the speed of light in water emit Cherenkov photons, which are subsequently detected by photomultiplier tubes (PMTs) lining the detector volume. However, the number of detectable Cherenkov photons is often limited by the wavelength-dependent sensitivity of PMTs and the absorption properties of water. Improving the light yield and collection efficiency can directly enhance the detector performance, particularly for low-energy or rare-event signals.

The AMoRE-II (Advanced Mo-based Rare process Experiment) project aims to search for neutrinoless double beta decay using cryogenic detectors operated in a deep underground laboratory. To suppress muon-induced backgrounds, the experiment employs a large water Cherenkov detector as an active muon veto system, filled with about 60 tons of deionized water and instrumented with 48 PMTs to detect Cherenkov light produced by passing muons~\cite{AMoRE:2024tjb}. Enhancing the photon detection efficiency of this veto detector is essential for improving the background rejection power and ensuring the sensitivity goals of the experiment.

One approach to increase the number of detectable photons in water Cherenkov detectors is to use wavelength shifters (WLS) that absorb ultraviolet Cherenkov light and re-emit it at longer wavelengths, where PMTs are more efficient~\cite{Dai:2008cp}. Among various candidates, 4-methylumbelliferone (4-MU) is known for its strong absorption in the UV region and efficient fluorescence in the blue range, making it a suitable WLS additive in water-based systems~\cite{Sweany:2011qh}.

This study investigates the use of 4-MU as a wavelength-shifting additive to improve photon detection efficiency in water Cherenkov detectors. We tested a range of 4-MU concentrations in ultrapure water and measured the Cherenkov light yield using cosmic-ray muons as a natural source. The experimental procedure and results are presented in the following sections.

\section{Optical Properties of 4-MU}
\label{sec:Opts}
To evaluate the optical behavior of 4-MU~\cite{AlfaAesarA10337, ChemicalBook4MU, Cayman4MU} in water, we measured its absorbance and fluorescence emission spectra. These measurements are relevant for assessing the potential of 4-MU as a wavelength-shifting material in water Cherenkov detectors.

The samples were prepared by first dissolving 4-MU in ethanol to ensure complete dissolution, and then mixing the solution with ultrapure water to achieve the desired concentrations. This method allows 4-MU, which is insoluble in water, to be effectively loaded into the aqueous medium. Our results indicate that when ethanol is present in an amount sufficient to dissolve the target amount of 4-MU, the compound can be successfully mixed with water without visible precipitation.

\begin{figure}[htbp]
\centering
\includegraphics[width=0.55\textwidth]{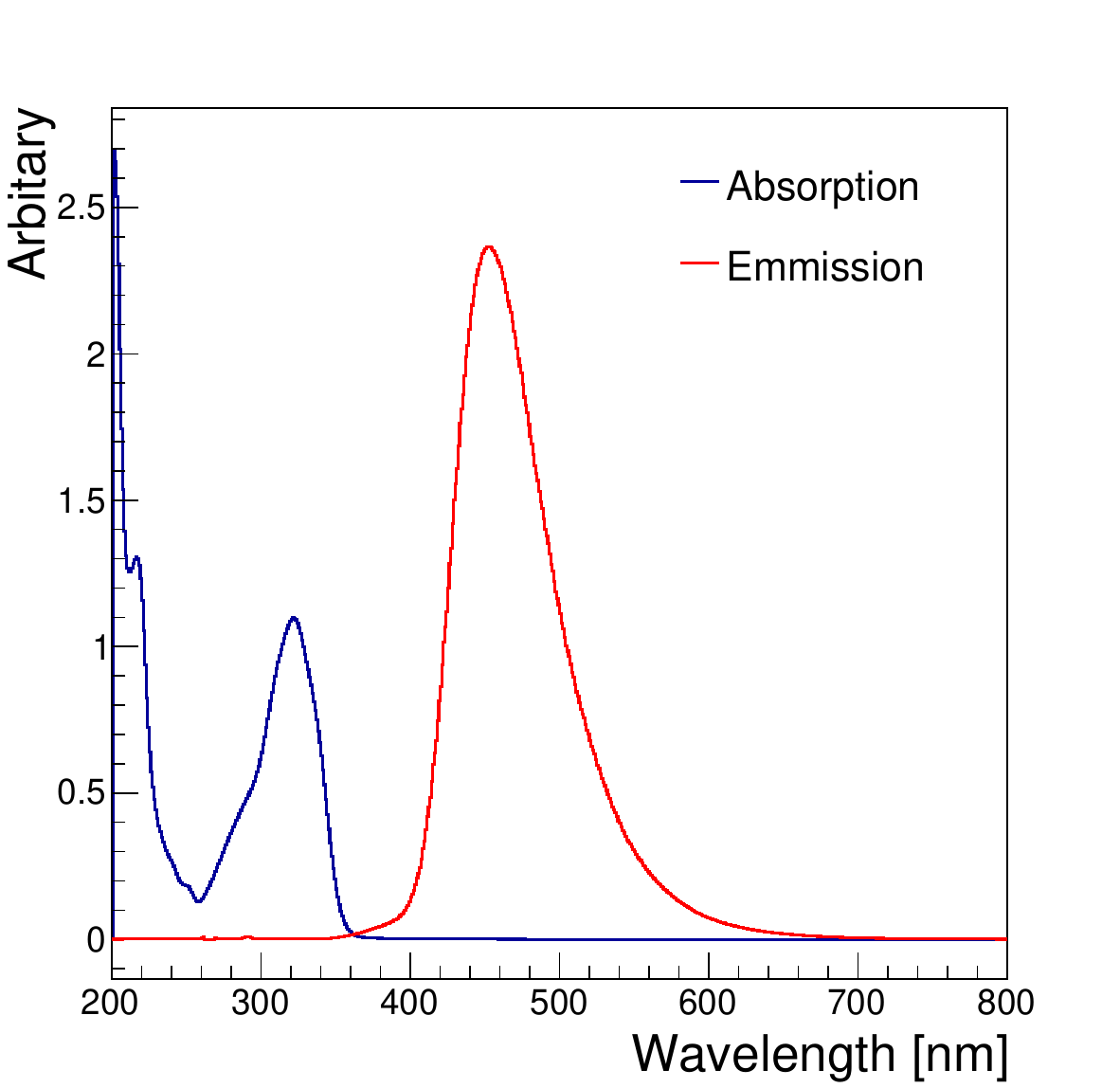} 
\caption{Absorbance and fluorescence emission spectra of 4-MU in water.
The absorbance spectrum (blue) shows strong ultraviolet absorption, while the fluorescence emission spectrum (red), excited at 320 nm, peaks in the blue region near 450 nm. These properties make 4-MU a suitable wavelength shifter for enhancing the sensitivity of bialkali PMTs.\label{fig:4mu_spectra}}
\end{figure}

The absorbance spectrum of 4-MU was measured using a UV-Vis spectrophotometer. It shows strong absorption in the ultraviolet region, with a clear peak near 320 nm. This corresponds to the wavelength range where Cherenkov radiation is most intense but PMT sensitivity is limited.
Fluorescence emission spectra were obtained using a fluorescence spectrometer, with the excitation wavelength set near the absorbance maximum. The emission spectrum is broad and centered around 450 nm, which overlaps well with the peak quantum efficiency region of typical bialkali PMTs.
Figure~\ref{fig:4mu_spectra} shows both the absorbance and emission spectra of 4-MU in water, highlighting the spectral shift between absorption and emission.
To check for possible self-absorption or fluorescence quenching, measurements were carried out at several concentrations, including 0.1, 0.5, and 1 ppm by mass. No significant spectral changes were observed within this range, indicating that 4-MU maintains stable optical properties at these low concentrations.

\begin{figure}[htbp]
\centering
\includegraphics[width=0.9\textwidth]{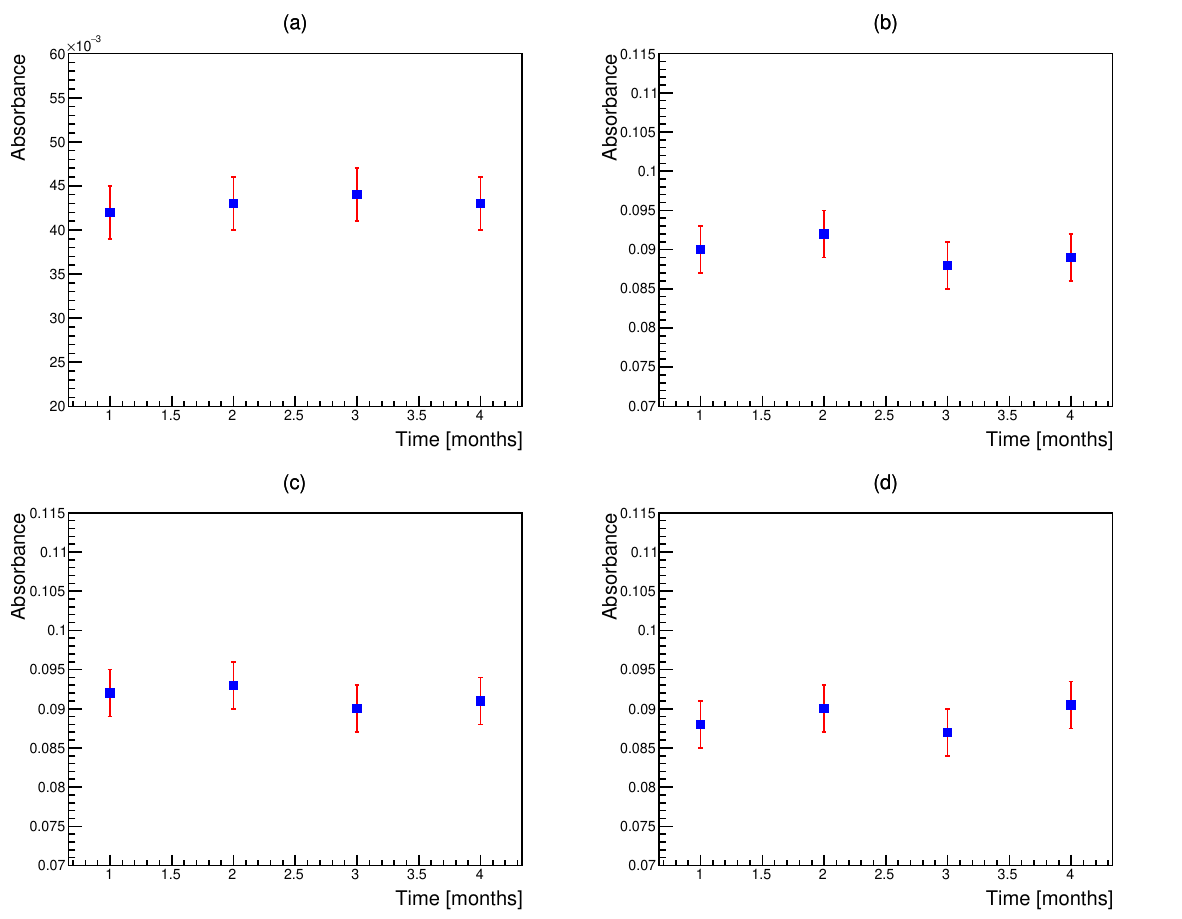} 
\caption{Time evolution of absorbance values for 4-MU solutions over a 4-month period: (a) 0.5 ppm of 4-MU, 0.1\% Ethanol, (b) 1 ppm of 4-MU, 0.1\% Ethanol, (c) 1 ppm of 4-MU, 0.5\% Ethanol, (d) 1 ppm of 4-MU, 1\% Ethanol \label{fig:jan_abs2}}
\end{figure}

To explore short-term stability, we monitored the absorbance spectra of samples prepared with various 4-MU concentrations (0.5 and 1 ppm by mass) over approximately four months. In this process, we also varied the amount of ethanol used to dissolve 4-MU prior to mixing with water. Figure~\ref{fig:jan_abs2} shows the absorbance at 420 nm, measured at multiple time points and under different ethanol and 4-MU concentrations, with no noticeable changes observed over the monitored period or due to ethanol variation. While these results indicate that 4-MU maintains stable optical properties in the short term, additional studies over longer timescales are needed to confirm its long-term stability for experimental applications.

\section{Light yield and muon detection efficiency in a prototype detector}
\label{sec:proto}
A cylindrical prototype water Cherenkov detector was constructed to evaluate the effect of 4-MU concentration on light yield and muon detection efficiency. The detector consists of a stainless steel cylinder, 70 cm in height and 40 cm in diameter, filled with either ultrapure water or 4-MU-loaded water, depending on the test conditions. A single 10-inch PMT (Hamamatsu R7081) is mounted at the top center inside the tank using a dedicated holder, with its photosensitive surface fully submerged to ensure efficient collection of Cherenkov photons. The inner surface of the cylinder is lined with Tyvek to enhance photon collection through diffuse reflection. The detector was operated in a dark environment to suppress ambient light contamination.

To provide a coincidence trigger for vertically traversing cosmic-ray muons, a pair of plastic scintillator counters was placed above and below the tank. Since their cross-sectional area was smaller than that of the tank, a clean selection of muon events passing through the detector volume was ensured without underestimating the efficiency.
The PMT signals were digitized using a flash analog-to-digital converter (FADC) developed by Notice Korea, featuring a sampling rate of 500 MS/s and a resolution of 12 bits. The digitized waveforms were stored for offline analysis, from which the total charge and peak amplitude were extracted to evaluate the light yield. A schematic and photographs of the detector and trigger configuration are shown in Figure~3.

\begin{figure}[htbp]
\centering
\includegraphics[width=0.8\textwidth]{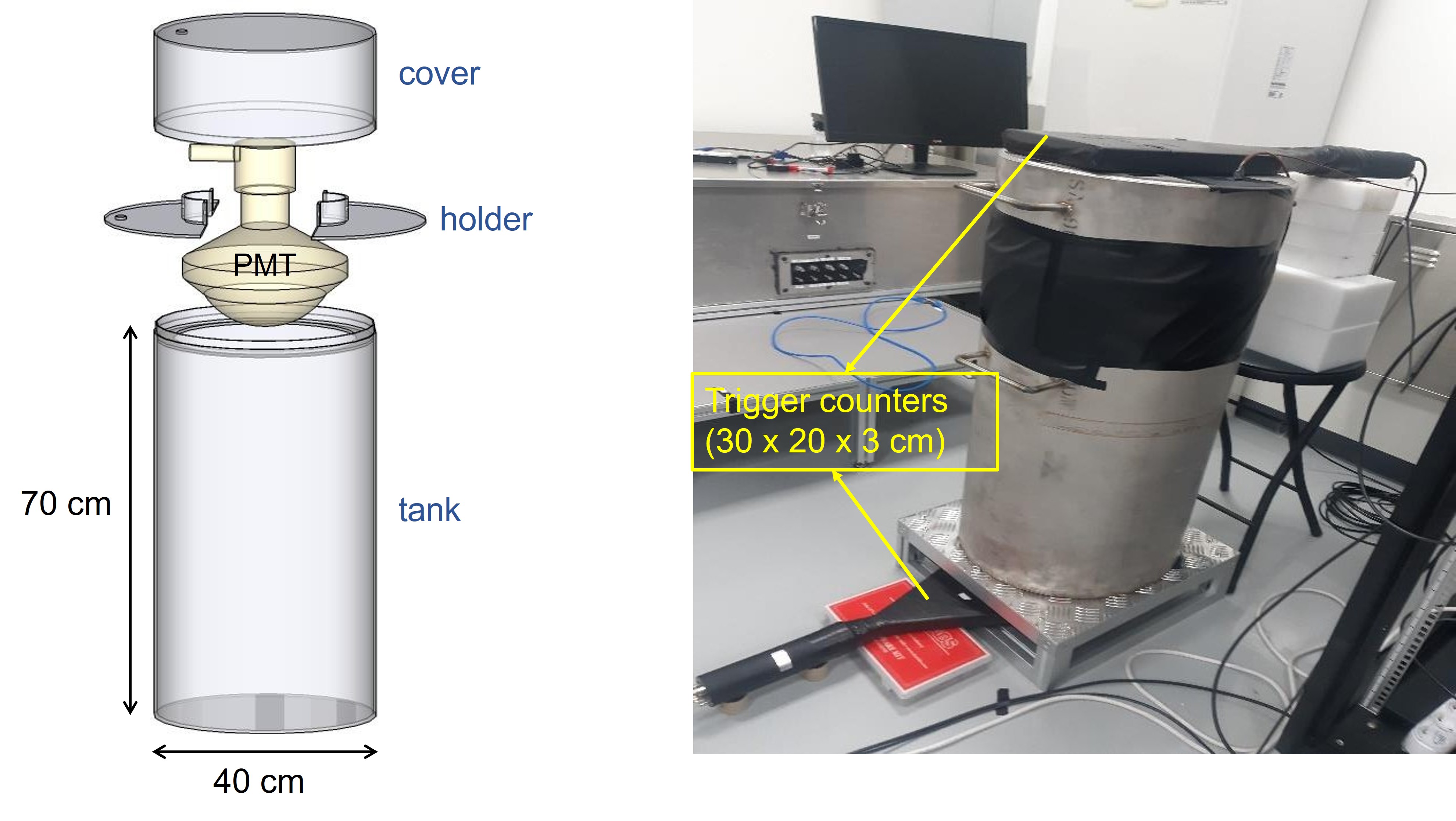} 
\caption{A schematic diagram of a prototype water Cherenkov detector (left) and a photo of experimental setup (right) \label{fig:jan_abs3}}
\end{figure}

To evaluate the light yield from cosmic-ray muons, we applied a selection based on the integrated charges from the upper and lower plastic scintillator detectors. Figure~4 (a) shows a two-dimensional scatter plot of the integrated charges recorded by the upper and lower trigger detectors. A rectangular selection region was defined to identify vertically traversing muons that deposit large amounts of energy in both detectors. Events falling within this region were classified as muon candidates passing through the detector volume.
Figure~4(b) presents the distribution of the number of photoelectrons (NPE) recorded by the PMT for all triggered events in ultrapure water. The inset shows the spectrum after applying the muon selection. The selected muon events exhibit a prominent peak, and the most probable value of the distribution was extracted to be $854.1 \pm 1.6$ NPEs from a Landau fit, which represents the characteristic light yield from cosmic-ray muons and serves as a reference for comparison with data taken using 4-MU-loaded water.

\begin{figure}[htbp]
\centering
\begin{subfigure}[b]{0.48\textwidth}
    \centering
    \includegraphics[height=7.cm]{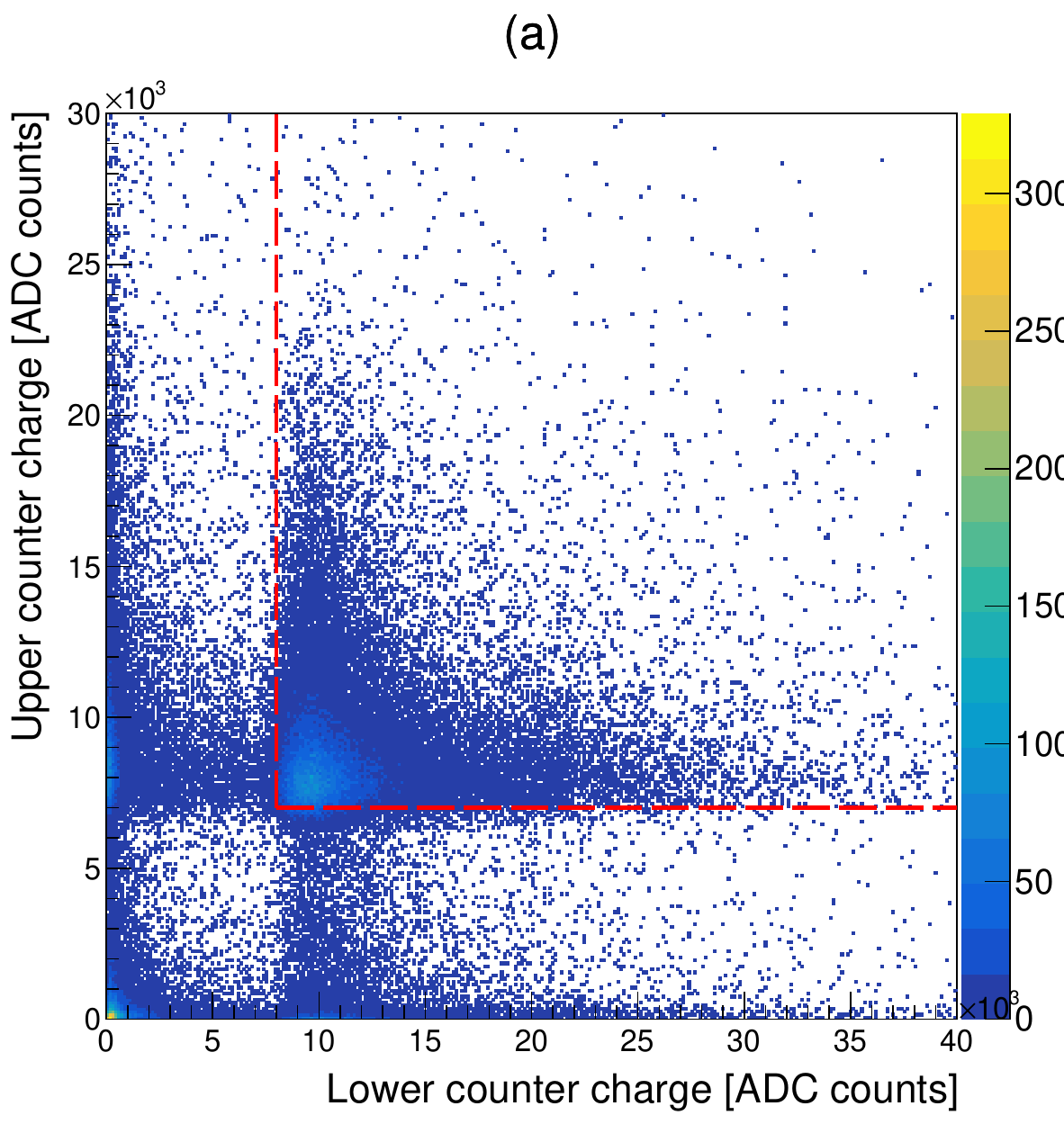}
\end{subfigure}
\hfill
\begin{subfigure}[b]{0.48\textwidth}
    \centering
    \includegraphics[height=7.cm]{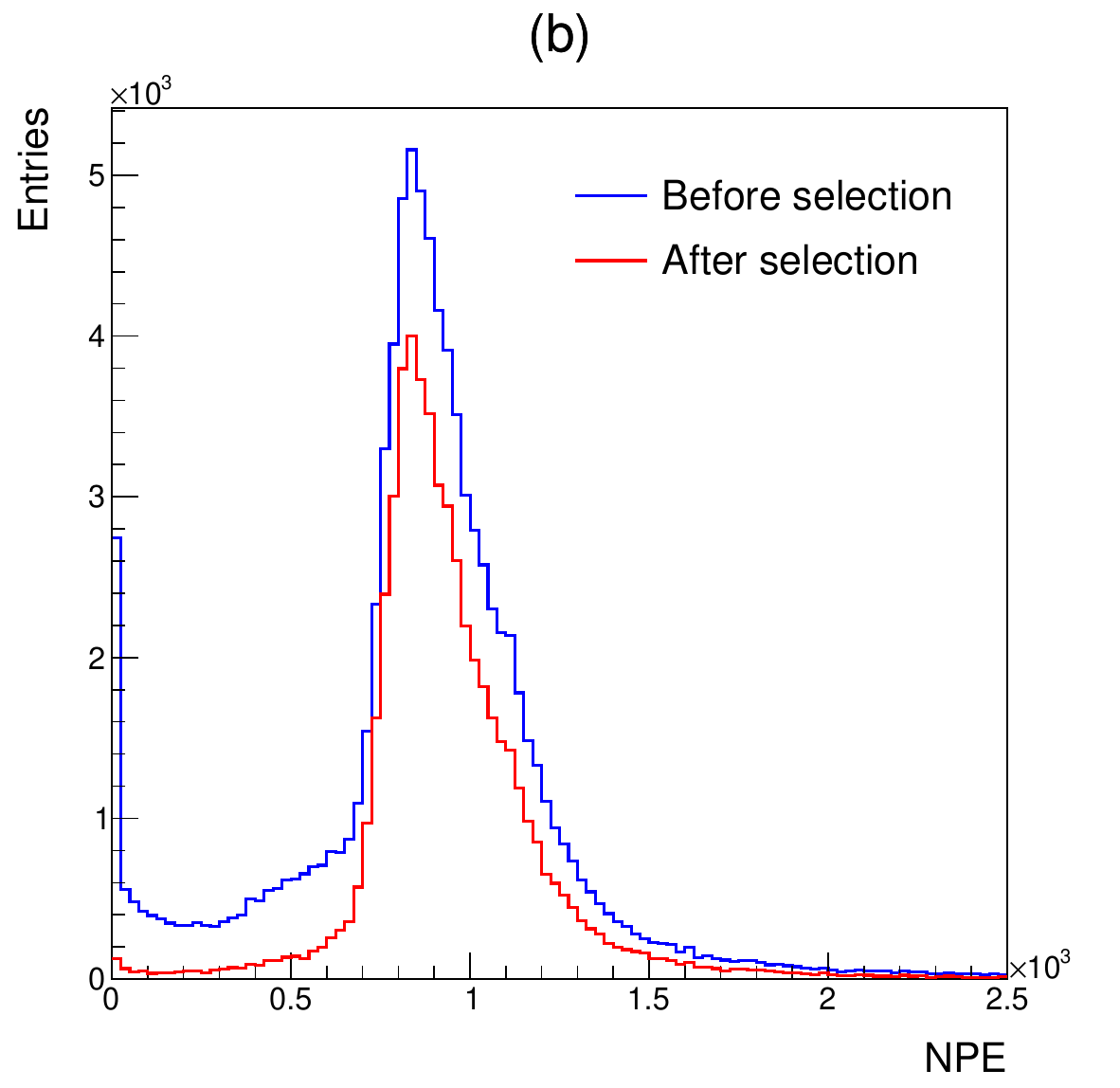}
\end{subfigure}
\caption{(a) Charge correlation between upper and lower scintillator counters used for muon tagging. The red dashed lines indicate the selection cuts for through-going muon events. (b) NPE distribution obtained in pure water, shown before (blue) and after (red) applying the muon selection criteria.}
\label{fig:combined}
\end{figure}

\begin{figure}[htbp]
\centering
\begin{subfigure}[b]{0.48\textwidth}
    \includegraphics[width=\textwidth]{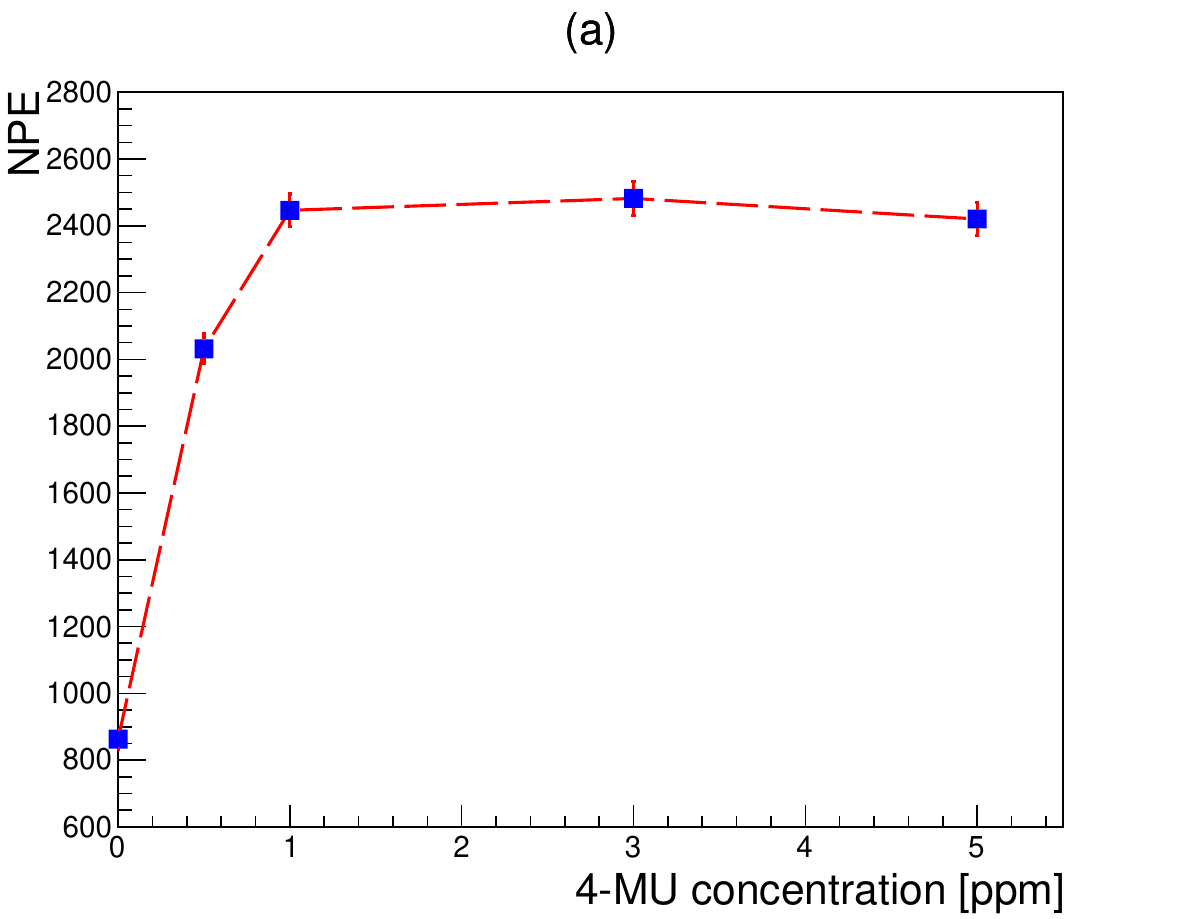}
\end{subfigure}
\hfill
\begin{subfigure}[b]{0.48\textwidth}
    \includegraphics[width=\textwidth]{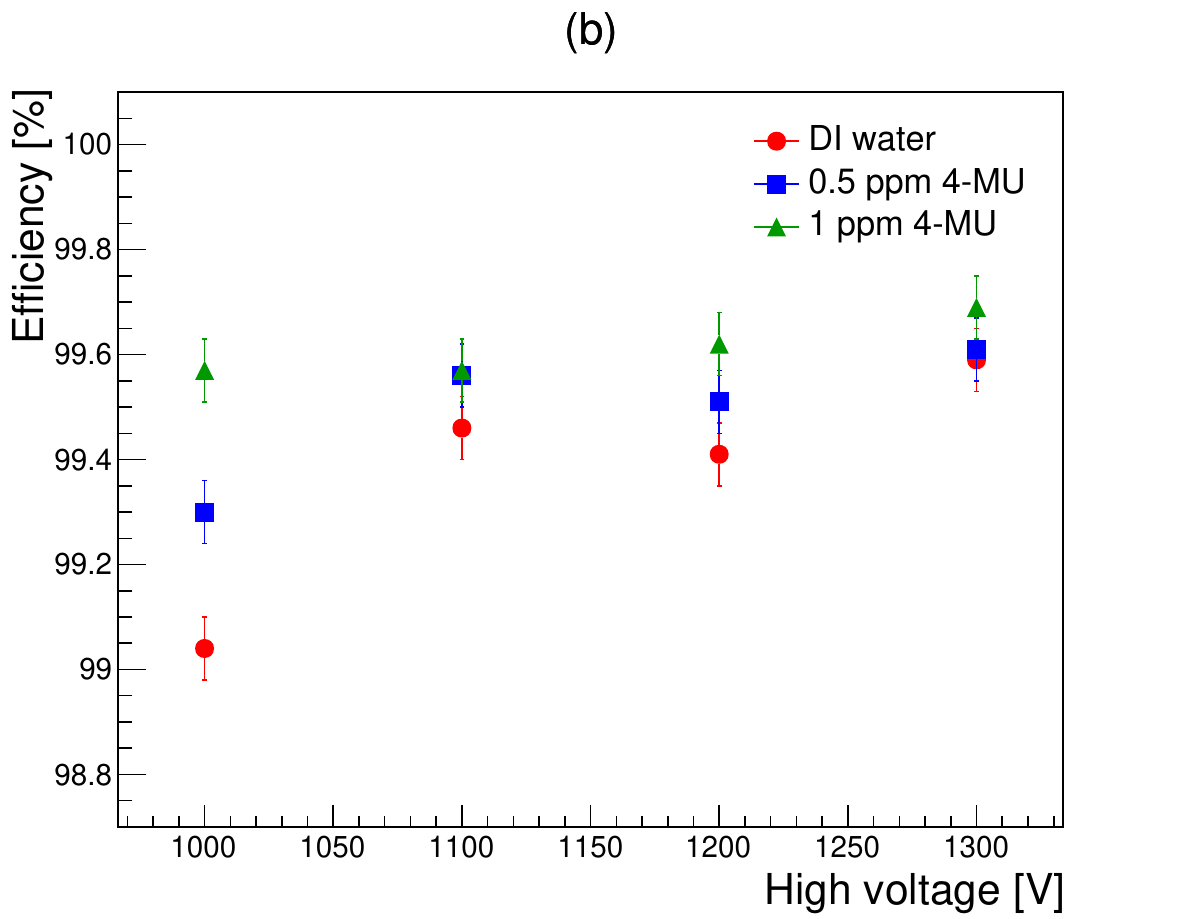}
\end{subfigure}
\caption{(a) Average NPE detected from muon events as a function of 4-MU concentration in water. A rapid increase in light yield is observed up to 1 ppm, beyond which the performance saturates. (b) Muon detection efficiency as a function of PMT high voltage for three different conditions: DI water, 0.5 ppm 4-MU, and 1 ppm 4-MU.}
\label{fig:deteff}
\end{figure}

The dependence of light yield on 4-MU concentration was investigated by preparing a series of water samples with increasing levels of 4-MU doping in 1\% ethanol. Figure~\ref{fig:deteff} (a) shows the most probable value of NPE measured for each concentration. A clear increase in light yield is observed as the 4-MU concentration increases from 0 to 1 ppm, with the NPE nearly tripling compared to that of pure water. Beyond 1 ppm, the gain in light yield begins to saturate, showing little improvement between 1 and 5 ppm. This behavior suggests that 1 ppm is close to the optimal concentration for maximizing Cherenkov light collection under the given detector conditions.
Figure~\ref{fig:deteff} (b) shows the muon detection efficiency measured at various PMT high voltage (HV) settings for three representative samples: pure water, 0.5 ppm 4-MU, and 1 ppm 4-MU. The efficiency is defined as the fraction of trigger events in which the PMT signal exceeded a predefined threshold. The results show a small but noticeable improvement in detection efficiency with increasing 4-MU concentration, particularly at lower HV settings. At 1 ppm, the detection efficiency reaches approximately 99.7\% across the tested HV range.

\begin{figure}[htbp]
\centering
\includegraphics[width=0.7\textwidth]{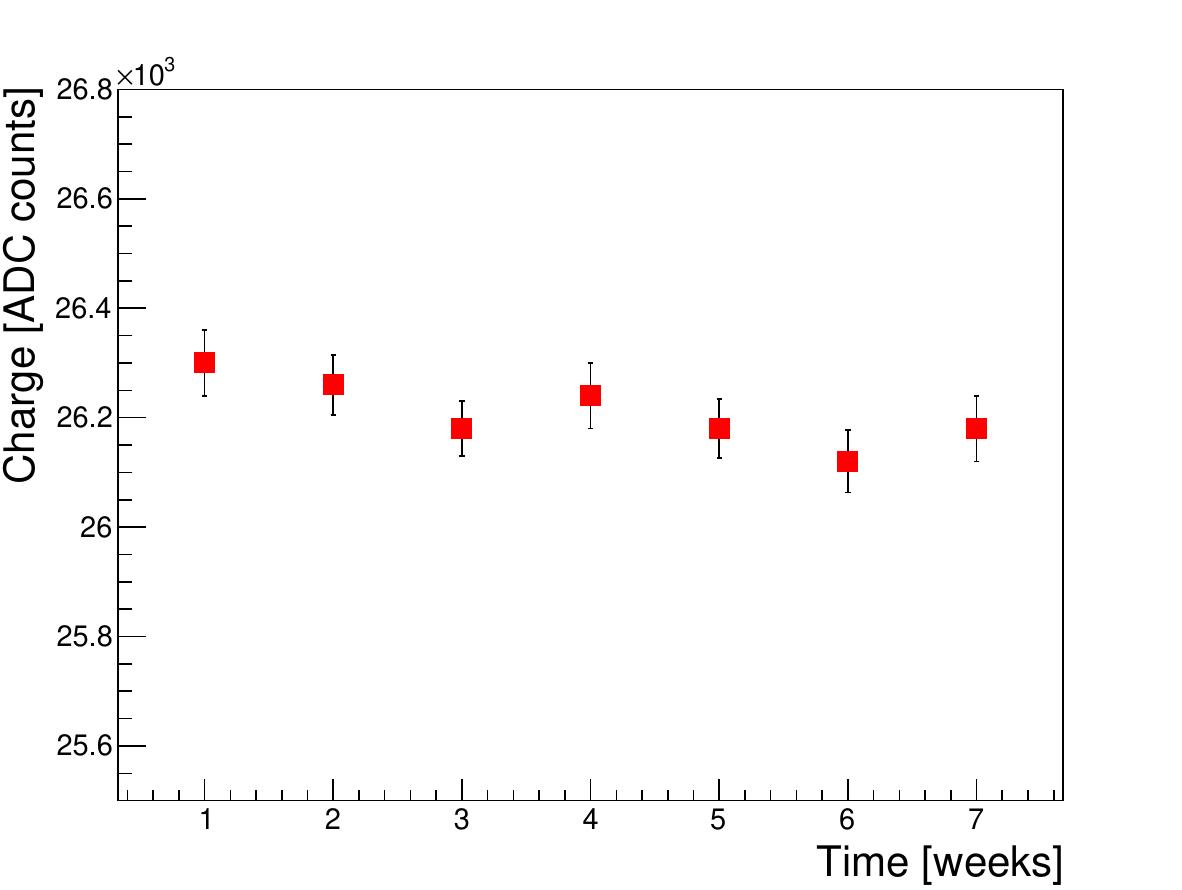} 
\caption{Average charge collected from muon events over time in a 1 ppm 4-MU solution. The charge remains stable, indicating no noticeable degradation in light yield. \label{fig:jan_abs}}
\end{figure}

To evaluate the temporal stability of 4-MU in water, we monitored the light yield of a 1 ppm solution over a period of approximately seven weeks. The charge collected from muon events was used to monitor the light yield, with measurements performed weekly under consistent conditions. As shown in Figure~\ref{fig:jan_abs}, the measured charge values fluctuate within a narrow range, with no evident decreasing trend. Although the measurement period is not sufficient to assess long-term chemical stability, the results indicate that the light yield remains stable over several weeks.

\section{Conclusion}
\label{sec:conc}
We have investigated the applicability of 4-methylumbelliferone (4-MU) as a wavelength-shifting additive for enhancing the performance of water Cherenkov detectors. Optical measurements confirmed that 4-MU efficiently absorbs ultraviolet light and re-emits it in the blue region, which aligns well with the sensitivity of typical bialkali PMTs. Light yield measurements using a prototype detector demonstrated that even small amounts of 4-MU significantly enhance the Cherenkov photon yield. At a concentration of 1 ppm, the light yield increased by approximately a factor of three compared to pure water, with a corresponding improvement in muon detection efficiency, particularly at lower high-voltage settings. The performance was observed to saturate around 1 ppm, suggesting an optimal concentration for practical application.

Furthermore, no signs of chemical or optical degradation were observed over a monitoring period of seven weeks, indicating that 4-MU can maintain stable light yield over relevant experimental timescales. These results demonstrate that 4-MU is a promising and practical additive for improving the photon collection efficiency of water Cherenkov detectors.

\acknowledgments
This research was supported by the Institute for Basic Science under project code IBS-R016-A1 Republic of Korea, the National Research Foundation of Korea (NRF) grant funded by the Korean government (MSIT) (NRF-2021R1A2C1013761).


\bibliographystyle{JHEP}
\bibliography{biblio.bib}

\end{document}